# Minimal Switch Step Tracking Control of Switched Systems with Application to Induction Motor Control

Babak Tavassoli

*Abstract*— The problem of step tracking control with a switching input and without any continuous-valued inputs is considered. The control objective is to reduce the number of switchings to a minimal value. This approach finds interesting applications when switching comprises costs and should be avoided. To solve the problem, a state dependent switching strategy should be designed and the resulting closed loop is indeed a hybrid system. Therefore, first we investigate the conditions on a hybrid system for being the desired solution. Then, we propose a method for designing the switching strategy such that the closed loop as a hybrid system solves the problem. The proposed method is applied to the induction motor control problem which results in relatively simple and efficient control algorithm. Comparison with the direct torque control for induction motors show that our method has a superior performance in reducing the number of mode switches.

## I. Introduction

Development of technologies in the recent decades has led to emergence of new control applications. In a class of such applications which is usually referred to as switched systems, the dynamical model of system has a discrete-valued mode [1, 2, 3]. The discrete mode affects the continuous-valued state variable dynamics. If the discrete mode also depends on the continuous state variables, then we have a hybrid system [4, 5]. Switched and hybrid systems find important applications in many areas such as mechanical systems, process control, automotives, aerial and ground transport systems, power systems, networked control systems and etc. An important problem in this domain is when the discrete mode acts as an input of the system which is referred to as the switching control problem or designed switching [3, 1]. A category of applications for this case are power electronic circuits and drives with electronic switches that have attracted applications of the switched and hybrid systems theories in various forms [6, 7, 8, 9, 10]. However, the discrete mode may have other roles such as being a source of randomness in the system [11] or representing the structure of a complex system [12]. Various control problems have been studied for the cases of switched and hybrid systems that include the problem of tracking a reference output [13, 14, 15, 16].

In this work, we consider the step tracking control problem for switched and hybrid systems using the discrete mode as an input of the system. The main control objective is to track a step function as the desired output. Our second objective is to apply a minimal number of mode switches in order to achieve the tracking objective. This approach to the problem has not been considered previously. In the existing results, either a continuous-valued input is available [13, 14] or the number of switches is not an issue and only special cases of reference outputs are considered [15, 16]. Our problem finds important applications when there is a cost associated with each mode switch which motivates the reduction of the total number of mode switches. An example is power electronic circuits with switching elements in which an amount of energy loss is associated with each switching [8, 7]. To solve the problem, we observe that the closed loop resulting from the switching control is in fact a hybrid dynamical system. Therefore, first we formulate a hybrid system which is able to track a step command. After this step, we design the switching such that the closed loop hybrid system solves the problem. We also provide the solvability conditions. The switching is designed in a way that the time between two successive mode switches is maximized. This does not necessarily result in the minimum number of switchings over time, but the achived number of switchings can be regarded as minimal (suboptimal). However, this approach simplifies the control algorithm and

.

reduces its computations significantly, in comparison with optimization based methods such as the finite horizon optimal control approach in [8]. We will apply our method to an induction motor. The result is presented as a control algorithm which is ready for implementation. Then, we make a performance comparison with the direct torque control method (DTC) [17, 18]. This method is widely regarded as the successful control method for induction motors. But, this method suffers from requiring high switching frequencies for reducing the amplitude of the tracking error fluctuations. Simulations are provided that show our method applies a considerably smaller number of switches in comparison with the DTC method.

The organization of the paper is as following. Problem formulation for hybrid and switched systems is performed in section two. Solution of the problem together with the conditions for solvability is provided in section three. The results are applied to the induction motor in section four where a comparison is also made with the DTC method. Conclusions are made at the end.

*Notation*: In the following, $\mathbb{R}$ is the set of real numbers, $\mathbb{R}_+$ is the set of non-negative real numbers and $\mathbb{Z}_+$ is the set of non-negative integers. The Euclidian norm of a vector $\xi \in \mathbb{R}^n$ is denoted by $\|\xi\|$ and its $i$th element is denoted as $[\xi]_i$. The boundary of a set $M$ in a metric space is denoted by $\partial M$ and its closure is denoted by cl[$M$]. For a set $A$, its cardinality is denoted by $|A|$ and the set of all subsets of $A$ is denoted by $2^A$. We say that a mapping $\Phi:\mathbb{R}^n \times \mathbb{R} \to \mathbb{R}^n$ is a transition function for a smooth vector field $f:\mathbb{R}^n \to \mathbb{R}^n$, if the differential equation $\partial \Phi(\xi,t)/\partial t = f(\Phi(\xi,t))$ with $\Phi(\xi,0) = \xi$ is satisfied for every $\xi \in \mathbb{R}^n$ and $t \in \mathbb{R}$.

## II. PROBLEM FORMULATION

In this section, the tracking problem using a switching input is formulated after some preliminaries.

### A. Switched and Hybrid Systems

A hybrid system has a set of continuous state variables and a discrete state variable that interact with each other while evolving along time. A change of the discrete state is referred to as a jump. Between jumps, the discrete state is constant and the vector of continuous states evolves according to an ordinary differential equation (ODE) which depends on the discrete state. This type of evolution of the state is denoted as a flow. In a general hybrid system, the continuous states may also change value at a jump which is denoted as a reset. However, for simplicity we ignore the reset function and rewrite the definition of hybrid system in [4] as below.

*Definition 1*: A Non-Reset Hybrid System (NRHS) is a sextuple $H = (X, U, E, \{\text{In}_x\}_{x \in X}, \{f_x\}_{x \in X}, \{\text{Gu}_t\}_{t \in E})$ with

- $X$ is a finite set of discrete states;
- a set of inputs $U$;
- a transition relation $E \subseteq X \times U \times X$;
- a non-empty set $\text{In}_x \subseteq \mathbb{R}^n$ for each $x \in X$ denoted as invariant set of $x$;
- a smooth vector field $f_x : \text{In}_x \to \mathbb{R}^n$ for each $x \in X$;
- a guard set $\emptyset \neq \text{Gu}_{(x,u,x')} \subseteq \text{In}_x$ for each $(x,u,x') \in E$.

In the above definition, $n$ is the dimension of the continuous state. The state of the NRHS is $(x, \xi)$ with $x \in X$ and $\xi \in \mathbb{R}^n$.

*Remark 2*: If $\xi$ belongs to the interior of $\text{In}_x$ and it also belongs to $\text{Gu}_t$ for some $t = (x,u,x')$, then both jump and flow are possible at $(x, \xi)$. This situation is regarded as a form of uncertainty [5]. In this paper we avoid such an uncertainty by assuming that a jump has priority over flow (i.e. when both are possible a jump occurs). Another form of uncertainty is possibility of having two jumps with the same input which is also avoided by the following assumption.

*Assumption 3*: For every $(x,u,x')$, $(\bar{x},\bar{u},\bar{x}') \in E$ we have

$$u = \bar{u} \ \wedge x = \bar{x} \ \Rightarrow \ x' = \bar{x}'$$

A property of the Definition 1 (inherited from [4]) is that the input acts only on jumps without affecting the flows. This property conforms to our objective to control with only switchings.

A switched system is an NRHS with $\text{In}_x = \mathbb{R}^n$, $\text{Gu}_t = \mathbb{R}^n$ for every $x \in X$, $t \in E$, $U = X$ and $E = \{(x,u,x') \in X \times U \times X : u = x'\}$. Hence, we have a more compact definition for a switched system as the following.

*Definition 4*: A switched system is a pair $S = (X, \{f_x\}_{x \in X})$ composed of a finite set of discrete states $X$ and a set of vector fields $f_x : \mathbb{R}^n \to \mathbb{R}^n$ for each $x \in X$.

According to $U = X$ when representing a switched system as an NRHS, $x$ acts as an input of the switched system. Hence, in the case of a switched system the discrete state $x \in X$ may be referred to as switching input or *mode*.

The switched system of Definition 4 is a special case of the NRHS in Definition 1 from a mathematical point of view. However, we can build an NRHS from a switched system by selecting guards and invariant sets to restrict the switchings or the jumps. Hence, a switched system may be regarded as more general than an NRHS from a practical viewpoint. Based on this observation, we add the following definition.

*Definition 5*: An NRHS $H = (X, U, E, \{\text{In}_x\}_{x \in X}, \{f_x\}_{x \in X}, \{\text{Gu}_t\}_{t \in E})$ is a restriction of a switched system $S = (X, \{f'_x\}_{x \in X})$ if $U = X$, $E = \{(x,x',x') : x,x' \in X\}$ and $f'_x = f_x$ for every $x \in X$.

For an NRHS which is a restriction of a switched system we can compactly write the triple $(x,x',x') \in E$ as an ordered pair $(x,x')$.

### B. Problem statement

To define the tracking problem we need to define an output for the system. Since the discrete variable $x \in X$ acts as an input of the switched system, we define the output as a function of the continuous state variables only.

*Definition 6*: For an NRHS or for a switched system with continuous state vector $\xi \in \mathbb{R}^n$, the vector $y \in \mathbb{R}^m$ is an output if there exist a mapping $h : \mathbb{R}^n \to \mathbb{R}^m$ denoted as the output function such that $y = h(\xi)$ for every $\xi \in \mathbb{R}^n$.

The output at time $t \in \mathbb{R}_+$ denoted by $y(t)$ is required to track a constant desired output $y_d$ which belongs to a set $Y_d \subseteq \mathbb{R}^m$ as in the following.

*Problem 7*: Consider the NRHS in Definition 1 with the set of inputs $U$, an output whose value at $t$ is denoted by $y(t) \in \mathbb{R}^m$, and a vector of tracking error bounds $\varepsilon \in \mathbb{R}_+^m$. At a jump instant $t'$, select $u(t') \in U$ according to the hybrid state of NRHS at $t'$ and the desired output $y_d \in Y_d$ such that for every $t \in [t',t'']$ and $1 \le i \le m$ the condition in (1) holds where $t''$ is either the time instant of the next jump or $t''$ is infinite if there are no jumps after $t'$.

$$|[y(t) - y_d]_i| < [\varepsilon]_i \quad \vee \quad \frac{d}{dt}|[y(t) - y_d]_i| < 0 \tag{1}$$

The above conditions requires that each element of the tracking error $y(t) - y_d$ is either within the error bound or it moves toward such a region.

If at every jump instant it is possible to select the input such that (1) is satisfied, then we say that Problem 7 is solvable. If the Problem 7 is solvable, then there may exist multiple choices of input that satisfy (1). In this case, we can select the input that optimizes a measure. In this work, we try to reduce the number of jumps (or the switchings) as below.

*Problem 8*: Considering the NRHS in Problem 7, among the multiple choices of $u(t') \in U$ at a jump instant $t'$ that result in establishment of (1) until the time instant of the next jump $t''$, select the one that maximizes $t'' - t'$.

For solvability of the Problem 8 there must be at least one choice of input that can establish (1) for $t \in [t', t'']$. Hence, the Problem 8 is solvable if and only if the Problem 7 is solvable. In the case of switched systems, there is an additional degree of freedom of choosing guards and invariant set to achieve the control performance. Therefore, the tracking control problem for switched systems is defined based on the Problem 8 as below.

*Problem 9*: For a switched system, find a restriction $H$ such that Problem 8 is solvable for $H$.

### III. MAIN RESULTS

Our main objective is to solve the Problem 9 which requires the solution of the Problems 7 and 8 for the NRHS case. In the context of the Definitions 1 and 4, for every $x \in X$, we denote the transition function of $f_x$ by $\Phi_x$. Also, we hide the dependencies on the desired output $y_d$ in (1) for brevity since it is a constant.

*A. Solution for the NRHS Case*

If the Problem 7 (or 8) is solvable, obtaining the solution is straightforward. Denoting the NRHS by $H$, first we define the set of hybrid states that belong at least to one guard set as below.

$$G(H) = \{(x, \xi) \mid x \in X, \xi \in \text{In}_x, \exists\, T = (x, u, x') \in E : \xi \in \text{Gu}_T\} \tag{2}$$

If a flow reaches $G(H)$, then a jump occurs according to the priority of jump over flow as described in Remark 2. We denote the time to next jump when starting to flow from a hybrid state $(x, \xi)$ by $\theta(x, \xi)$ which can be calculated as

$$\theta(x, \xi) = \sup\{t \in \mathbb{R}_+ \mid \forall s \in [0, t) : \Phi_x(\xi, s) \notin G(H)\}. \tag{3}$$

Then, for $\xi \in \mathbb{R}^n$ and the vector of error bounds $\varepsilon \in \mathbb{R}^m_+$, we denote by $X_\varepsilon(\xi)$ the set of modes in $X$ that can result in the establishment of (1) at a time instant at which continuous state is $\xi$ as below.

$$X_\varepsilon(\xi) = \Big\{x \in X \,\Big|\, \forall\, 1 \leq i \leq m\ |[y(0) - y_d]_i| < [\varepsilon]_i \vee \frac{d}{dt}|[y(t) - y_d]_i|\big|_{t=0} < 0,\ y(t) = h(\Phi_x(\xi, t))\Big\}$$

The time derivative in the above equation can be eliminated by applying the chain rule to obtain

$$X_\varepsilon(\xi) = \Big\{x \in X \,\Big|\, \forall\, 1 \leq i \leq m\ |[h(\xi) - y_d]_i| < [\varepsilon]_i \vee \text{sign}([h(\xi) - y_d]_i)\,[(\partial_\xi h(\xi)) f_x(\xi)]_i < 0\Big\} \tag{4}$$

We are interested in the set of modes that can keep (1) to hold until the next jump which is given by

$$X^c_\varepsilon(\xi) = \{x \in X \mid x \in X_\varepsilon(\Phi_x(\xi, t))\ \forall t \in [0, \theta(x, \xi))\} \tag{5}$$

We calculate the set of inputs that can cause a jump that result in the establishment of (1) as below.

$$U_\varepsilon(x, \xi) = \{u \in U \mid \exists\, x' \in X^c_\varepsilon(\xi) : (x, u, x') \in E, \xi \in \text{Gu}_{(x, u, x')}\} \tag{6}$$

At the end of each flow, when $G(H)$ is reached and a jump is about to occur, $U_\varepsilon(x, \xi)$ must be non-empty such that the establishment of (1) can continue in time. This gives the condition for solvability of the Problem 7 as

$$U_\varepsilon(x, \xi) \neq \emptyset \quad \forall\, (x, \xi) \in G(H). \tag{7}$$

If the Problem 7 is solvable, then the Problem 8 is also solvable since we should only select the element of $U_\varepsilon(x, \xi)$ which gives the largest value of $\theta(x, \xi)$ in (3).

We summarize this part as below.

*Proposition 10*: For an NRHS denoted by $H$, at every hybrid state $(x,\xi) \in G(H)$, the solution of Problem 7 is an arbitrary element from $U_\varepsilon(x,\xi)$ in (6) and the solution of Problem 8 is an element from $U_\varepsilon^*(x,\xi)$ in (8) with $\theta(x,\xi)$ in (3).

$$U_\varepsilon^*(x,\xi) = \{u \in U \mid u \in U_\varepsilon(x,\xi),$$
$$\forall (x,u,x'), (x,u',x'') \in E : \theta(x',\xi) \geq \theta(x'',\xi)\} \tag{8}$$

### B. Solution for the Switched System Case

Considering a switched system $S = (X, \{f_x\}_{x \in X})$, we should determine guard sets and invariant sets such that the Problem 7 (and 8) is solvable for the restriction of the switched system $S$. The invariant set $\text{In}_x$ for every $x \in X$ should be such that (1) holds in $\text{In}_x$. Therefore, based on the definition of $X_\varepsilon(\xi)$ in (4), for every $x \in X$ the largest possible invariant set is obtained as

$$\text{In}_x = \text{cl}[\{\xi \in \mathbb{R}^n \mid x \in X_\varepsilon(\xi)\}]. \tag{9}$$

We select the invariant sets to be the largest possible one in order to be able to have longer time intervals between jumps in the context of Problem 9. Also, to maximize controllability, we consider all possible jumps such that (1) remains valid and obtain the guard sets as

$$\text{Gu}_{(x,x')} = \{\xi \in \mathbb{R}^n \mid x \notin X_\varepsilon(\xi), x' \in X_\varepsilon(\xi)\}. \tag{10}$$

It is mentioned that at a point $\xi \in \partial \text{In}_x$ we may have $x \notin X_\varepsilon(\xi)$ due to taking the closure in (9). The invariant sets are defined to be closed sets to have intersection with the guards in (10) such that every flow in $\text{In}_x$ can be followed by a jump after arriving at $\partial \text{In}_x$.

By selecting the guards as in (10) we have $X_\varepsilon^c(\xi) = X_\varepsilon(\xi)$ for every $\xi \in \mathbb{R}^n$. In an NRHS which a restriction we have $U = X$ and according to (10), the equation (6) reduces to (11).

$$U_\varepsilon(x,\xi) = X_\varepsilon(\xi) \setminus \{x\} \tag{11}$$

By the definition of $X_\varepsilon(\xi)$, the guards and invariant sets in (9) and (10) establish (1). More precisely, a flow continues in $\text{In}_x$ if $x \in X_\varepsilon(\xi)$ and a jump occurs if $x \notin X_\varepsilon(\xi)$. Therefore, the possibility of such a jump is the only condition for solvability of Problem 9 which is expressed as below.

$$X_\varepsilon(\xi) \neq \emptyset \quad \forall \xi \in \mathbb{R}^n. \tag{12}$$

Also, (8) is simplified as

$$U_\varepsilon^*(x,\xi) = \arg\max_{x' \in X_\varepsilon(\xi) \setminus \{x\}} \theta(x',\xi) \tag{13}$$

We can now summarize this part as the following.

*Proposition 11*: For a switched system $S$, the Problem 9 is solvable if (12) holds. The corresponding restriction of $S$ to an NRHS which solves the problem is obtained by the invariant sets in (9) and the guards in (10). Also, the control input at each jump is selected from $U_\varepsilon^*$ in (13).

*Remark 12*: The condition (12) for solvability of the tracking problem can be interpreted as having sufficient actuation for the plant (or the switched system) through the available modes or the switching input. It is noticed that the condition (12) also depends on the choice of the output of the system and we may need to define the output suitably in order to solve the tracking problem (see the next part).

*Remark 13*: We can obtain a condition which is independent of $y_d$ and is a sufficient condition for (12). This can simplify the task of determining the solvability of the Problem 9. To do this, we require that only the expression on the second line of (4) must hold for some $x \in X$. To make it independent of $y_d$ we should be able to have any combination of signs for the elements of $(\partial/\partial \xi\, h(\xi))\, f_x(\xi)$ by selecting $x$ appropriately. This can be expressed as (14) which is the desired sufficient condition.

$$\forall \xi \in \mathbb{R}^n, \mu \in \{-1,+1\}^m \ \exists x \in X \ \mu = \text{sign}[(\partial/\partial \xi\, h(\xi))\, f_x(\xi)] \tag{14}$$

*Remark 14*: It is known that the state of a system may diverge even if its output converges. The part of state that can diverge when the output is fixed is known as the zero dynamics of a system. For solving every tracking problem it is required that the zero dynamics are stable. We fulfill this requirement by assuming that there exist $\delta > 0$ such that for the vector of error bounds $\varepsilon$, every $y_d \in Y_d$ and every initial condition we have

$$\{\forall t > 0: |[y(t) - y_d]_i| < [\varepsilon]_i\} \ \Rightarrow \ \{\forall t > 0: |\xi(t)| < \delta\} \tag{15}$$

### C. Using Modified Outputs

Assume that $z(t) \in \mathbb{R}$ is an output of the system with output function $h_z: \mathbb{R}^n \to \mathbb{R}$ that should track a step command but condition (12) fails with $y = z$. However, it may be possible to solve the tracking problem for another choice of $y$. We simply consider building the output $y$ from $z$ as

$$y(t) = \sum_{i=0}^{q} a_i\, z^{(i)}(t) \tag{16}$$

where $z^{(i)}(t)$ is the $i$th time derivative of $z(t)$. Denoting by $G_{zy}$ the transfer function from $y$ to $z$ according to (16), it must have stable poles such that

$$\sum_{i=0}^{q} a_i \gamma^i = 0 \ \Rightarrow \ \text{Re}\{\gamma\} < 0 \tag{17}$$

because when $y$ converges due to the tracking control we must also have the convergence of $z$.

It remains to find the relation between tracking error bounds for $z$ and $y$ denoted by $\varepsilon_z$ and $\varepsilon_y$ respectively. Denoting by $\|G_{zy}\|_\infty$ the induced $\mathcal{L}_\infty$ norm of $G_{zy}$, the error in $z$ will be bounded by $\varepsilon_y \|G_{zy}\|_\infty$ which gives

$$\varepsilon_z = \varepsilon_y \|G_{zy}\|_\infty \tag{18}$$

where $\|G_{zy}\|_\infty$ is calculated in terms of $h_{zy}$ the impulse response of $G_{zy}$ as

$$\|G_{zy}\|_\infty = \int_0^\infty |h_{zy}(s)|\, ds \ . \tag{19}$$

If we have a vector of outputs, we may need to perform this modification for some or all of them with possibly different values of $q$ in (16) for each of them.

## IV. APPLICATION TO INDUCTION MOTOR CONTROL

In this section we apply the results of the previous section to control an induction motor and make a comparison with the DTC method for the induction motor. For simplicity, we assume that the model of the motor is known and a correct estimation of the state variables is available. The theory of induction motor in this section is based on [19].

TABLE I. STATOR VOLTAGE FOR DIFFERENT SWITCHING COMBINATIONS

| Mode | $v_d$ | $v_q$ |
|---|---|---|
| 1 | 0 | 0 |
| 2 | $V_{DC}$ | 0 |
| 3 | $V_{DC}/2$ | $\sqrt{3}\, V_{DC}/2$ |
| 4 | $-V_{DC}/2$ | $\sqrt{3}\, V_{DC}/2$ |
| 5 | $-V_{DC}$ | 0 |
| 6 | $-V_{DC}/2$ | $-\sqrt{3}\, V_{DC}/2$ |
| 7 | $V_{DC}/2$ | $-\sqrt{3}\, V_{DC}/2$ |

*A. Switched System Model of an Induction Motor*

The key variables in the model of an induction motor are rotor angular speed $\omega$, stator d-q axes voltages $v_{ds}$, $v_{qs}$, rotor d-q axes fluxes $\lambda_{dr}$, $\lambda_{qr}$, stator d-q axes fluxes $\lambda_{ds}$, $\lambda_{qs}$, rotor d-q axes currents $i_{dr}$, $i_{qr}$, stator d-q axes currents $i_{ds}$, $i_{qs}$ and the generated torque $\tau$. The linkage equations between fluxes and currents is given as

$$\lambda = L\,i \tag{20}$$

$$\lambda = \begin{bmatrix}\lambda_{ds}\\ \lambda_{qs}\\ \lambda_{dr}\\ \lambda_{qr}\end{bmatrix},\ i = \begin{bmatrix}i_{ds}\\ i_{qs}\\ i_{dr}\\ i_{qr}\end{bmatrix},\ L = \begin{bmatrix}L_s & 0 & L_m & 0\\ 0 & L_s & 0 & L_m\\ L_m & 0 & L_r & 0\\ 0 & L_m & 0 & L_r\end{bmatrix}$$

where $L_s$, $L_m$, $L_r$ are electromagnetic induction coefficients. Also, the voltage equations of the motor are

$$\dot{\lambda} = -Ri + \omega C\lambda + Bv_{dq} \tag{21}$$

$$R = \begin{bmatrix}R_s & 0 & 0 & 0\\ 0 & R_s & 0 & 0\\ 0 & 0 & R_r & 0\\ 0 & 0 & 0 & R_r\end{bmatrix},\ C = \begin{bmatrix}0 & 0 & 0 & 0\\ 0 & 0 & 0 & 0\\ 0 & 0 & 0 & -1\\ 0 & 0 & 1 & 0\end{bmatrix},\ B = \begin{bmatrix}1 & 0\\ 0 & 1\\ 0 & 0\\ 0 & 0\end{bmatrix}$$

in which $R_s$ and $R_r$ are the stator and rotor resistances respectively and $v_{dq} = [v_{ds}\ v_{qs}]^T$. Torque $\tau$ is given by

$$\begin{aligned}\tau &= \frac{3}{2}\frac{P}{2}(i_{qs}\lambda_{ds} - i_{ds}\lambda_{qs})\\ &= \frac{3P}{4}\lambda^T CL^{-1}\lambda\end{aligned} \tag{22}$$

where $P$ is the number of poles of the motor. The mechanical equation of the motor is written as

$$J\dot{\omega} = -b\omega + \tau - \tau_L \tag{23}$$

in which $J$ is the rotor moment of inertia, $b$ is a friction coefficient and $\tau_L$ is a constant load.

If the stator voltages are supplied by an inverter with a DC voltage $V_{DC}$, then the different switching combinations of the inverter give the 7 possible values of $v_{dq}$ given in Table I. Hence, the set of modes of the switched system for the induction motor is selected as $X = \{1,2,...,7\}$ and the value of $v_{dq}$ that corresponds to $x \in X$ is denoted as $v_{dq}(x)$. According to (21) and (23), the vector fields of the switched system $f_x$, $x \in X$ are given as in (24) with the vector of continuous states $\xi = [\omega\ \lambda^T]^T$.

$$f_x\left(\begin{bmatrix}\omega\\ \lambda\end{bmatrix}\right) = \begin{bmatrix}-bJ^{-1}\omega + \frac{1}{J}\left(\frac{3P}{4}\lambda^T CL^{-1}\lambda - \tau_L\right)\\ -RL^{-1}\lambda + \omega C\lambda + Bv_{dq}(x)\end{bmatrix} \tag{24}$$

Model parameters for the induction motor which is considered in this section are also given in Table II.

*B. Torque Tracking Control of an Induction Motor*

Regarding Remark 14, it is known that the zero dynamics of an induction motor are stable [20]. Hence, we can apply the Proposition 11. Similar to the DTC method, we select the torque $\tau$ as an output of the motor. It is usual to maintain the stator flux magnitude $\lambda_{sm} = (\lambda_{ds}^2 + \lambda_{qs}^2)^{-1/2} = \sqrt{\lambda^T BB^T \lambda}$ at a fixed level. Hence we select the controlled outputs as $y = [\tau\ \lambda_{sm}]^T$ with the output function in (25).

$$\begin{bmatrix} \tau \\ \lambda_{sm} \end{bmatrix} = h\left(\begin{bmatrix} \omega \\ \lambda \end{bmatrix}\right) = \begin{bmatrix} \frac{3P}{4} \lambda^T C L^{-1} \lambda \\ \sqrt{\lambda^T B B^T \lambda} \end{bmatrix} \tag{25}$$

To investigate the solvability of the Problem 9 for the induction motor we use condition (14) in Remark 13. The expression $\partial h/\partial \xi \, f_x(\xi)$ in (14) can be calculated for the induction motor as below with the output function $h$ given in (25) and $f_x$ given in (24).

$\partial h/\partial \xi \, f_x(\xi) =$

$$\begin{bmatrix} \frac{3P}{4} \lambda^T (CL^{-1} - L^{-1}C)\left(Bv_{dq}(x) - RL^{-1}\lambda + \omega C\lambda\right) \\ \left(\lambda^T B v_{dq}(x) - \lambda^T B B^T R L^{-1} \lambda\right)/\sqrt{\lambda^T B B^T \lambda} \end{bmatrix}$$

With the above equation and a few reordering of terms we can write (14) as the set of relations in (26) that for every $\mu_1,\mu_2 \in \{-1,+1\}$ must hold for some $x \in X$.

$$\beta_i^T v_{dq}(x) - \alpha_i > 0 \quad i \in \{1,2\} \tag{26.1}$$

$$\beta_1 = \mu_1 B^T M_1 \lambda \quad , \quad \beta_2 = \mu_2 B^T \lambda \tag{26.2}$$

$$\alpha_1 = \mu_1 \lambda^T M_1 M_2 \lambda \quad , \quad \alpha_2 = \mu_2 \lambda^T B B^T R L^{-1} \lambda \tag{26.3}$$

$$M_1 = CL^{-1} - L^{-1}C = \frac{L_m}{L_s L_r - L_m^2} \begin{bmatrix} 0 & 0 & 0 & -1 \\ 0 & 0 & 1 & 0 \\ 0 & 1 & 0 & 0 \\ -1 & 0 & 0 & 0 \end{bmatrix} \tag{26.4}$$

$$M_2 = RL^{-1} - \omega C \tag{26.5}$$

In the above relations $\beta_1, \beta_2 \in \mathbb{R}^2$ and $\alpha_1, \alpha_2 \in \mathbb{R}$. Since in practice the values $L_s$, $L_r$ and $L_m$ are very close to each other, it can be approximately assumed that

$$\lambda^T \approx [\lambda_s^T \; \lambda_s^T] \tag{27}$$

which together with (26.4) results in $\beta_1^T \beta_2 = 0$. It means that $\beta_1$ and $\beta_2$ are perpendicular in the two dimensional plane. According to Table I there always exist a $x^*$ such that $v_{dq}(x^*)$ lays between $\beta_1$ and $\beta_2$, and we can write (28) with $\sphericalangle \beta$ being the angle of $\beta \in \mathbb{R}^2$ as a vector in the plane.

$$\frac{\pi}{12} \le |\sphericalangle \beta_i - \sphericalangle v_{dq}(x^*)| \le \frac{5\pi}{12} \quad i \in \{1,2\} \tag{28}$$

We can write (26.1) for $i \in \{1,2\}$ as

$$\|\beta_i\| \|v_{dq}(x^*)\| \cos\left(\sphericalangle \beta_i - \sphericalangle v_{dq}(x^*)\right) > \alpha_i \tag{29}$$

According to Table I, we always have $\|v_{dq}(x^*)\| = V_{DC}$. Hence, according to (28) the relation (29) is satisfied if $\|\beta_i\| V_{DC} \cos(5\pi/12) > |\alpha_i|$ for $i \in \{1,2\}$ that can be combined as below according to the definitions in (26).

$$V_{DC} > \frac{1}{\cos(5\pi/12)} \max\left\{ \frac{\lambda^T [M_1 M_2] \lambda}{\|B^T M_1 \lambda\|}, \frac{\lambda^T [BB^T R L^{-1}] \lambda}{\|B^T \lambda\|} \right\} \tag{30}$$

The above condition can be checked by a search for the worst case of the lower bound of $V_{DC}$ over the region of admissible state variables. For $0 < \omega < 50$ rad/s, and $\|\lambda\| < 5$, we obtain $V_{DC} > 433$ which verifies our value $V_{DC} = 450$.

To apply Proposition 11, it is needed to compute $\theta(x,\xi)$ in (3). We approximate this value by assuming that the time derivative of output is constant between two successive jumps. This is a reasonable assumption for the induction motor since the time interval between switches is very small with respect to the motor dynamics. With this assumption and by observing that $\theta(x,\xi)$ is the time at which the expression in (1) becomes false we can approximate $\theta(x,\xi)$ as below.

$$\theta(x,\xi) = \min\{\theta_i\} \tag{31}$$

$$\theta_i = \begin{cases} -([y - y_d]_i - \text{sign}([\dot{y}]_i)[\epsilon]_i)/[\dot{y}]_i & [\dot{y}]_i \neq 0 \\ \infty & [\dot{y}]_i = 0 \end{cases}$$

The value of $y$ in the above equation is obtained according to the output function in (25) and the value of $\dot{y}$ is calculated as $\partial h/\partial \xi \, f_x(\xi)$. We can summarize the control algorithm for the induction motor as below.

*Algorithm 15*:
Input: current state $\xi$, current mode $x$
Output: next mode $x^+$
1: Compute $X_\epsilon(\xi)$ from (4)     // use (24) and (25).
2: if $x \in X_\epsilon(\xi)$ then
3:    $x^+ := x$
4: else
5:    $x^+ := \arg\max_{x' \in X_\epsilon(\xi)} \theta(x',\xi)$    // use (31)
6: end if

*Remark 22*: If we select the rotor speed $\omega$ as a controlled output instead of the torque $\tau$, the conditions (14) or (12) fail to hold. However, if we select for example $\omega + \dot{\omega}$ as a modified controlled output in the context of the part C in previous section, then (14) can be satisfied.

*C. Simulation and Comparison*

In this part we simulate the motor control system with Algorithm 15. We also simulate the well known direct torque control (DTC) algorithm for induction motors [17, 18] and make a performance comparison between the two algorithms. The tracking error bounds are selected as $\epsilon = [0.1 \; 0.01]^T$. The desired torque is set to 50 N.m and the desired stator flux magnitude is set to 2 Wb. For simplicity, we do not include an estimator in our simulation and assume that the state variables are available (an estimator is required in practice). Simulation is performed for 10000 sampling periods of length 0.5 μs (the total simulation time is 5 ms). To inspect the steady state performance, initial conditions are selected such that the initial outputs are close the desired outputs.

TABLE II.  MOTOR PARAMETERS

| *Coefficient* | *Value* | *Units* |
|---|---|---|
| $L_s$ | 0.5676 | H |
| $L_r$ | 0.5676 | H |
| $L_m$ | 0.55 | H |
| $R_s$ | 1.19 | Ω |
| $R_r$ | 1.04 | Ω |
| $P$ | 4 | - |
| $J$ | 0.04 | kg. m$^2$ |
| $b$ | 0.07 | N. m. s |
| $\tau_L$ | 5 | N. m |
| $V_{DC}$ | 225 | V |

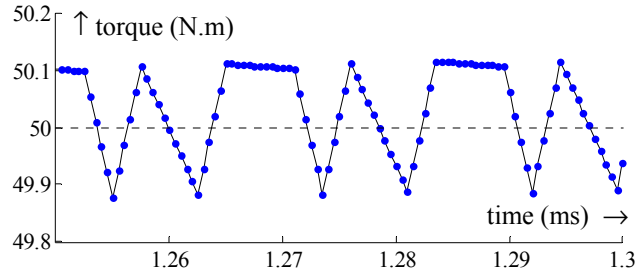

Figure 1. Operation of Algorithm 15.

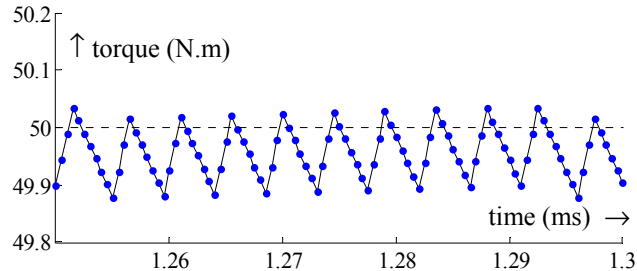

Figure 2. Operation of the DTC algorithm.

The number of mode switches for the DTC algorithm was obtained as 2435 during the simulation time. For Algorithm 15, the number of mode switches was obtained as 1308 during the same simulation time which is considerably smaller than the DTC mode switches. To see the difference in the operation of the two algorithms, small intervals of the two simulations are plotted in Figures 1 and 2 for comparison. In the figures, the desired torque is plotted as a dotted line. Time instants of execution of the control algorithm are highlighted by solid points on the curves. As it is seen Algorithm 15 is able to find a switch with a smaller slope to increase the time between mode switches of the inverter (or jumps of the switched system). Another point is that the tracking error of DTC is not bidirectional and tends to the negative values (which is related to the details of the DTC method).

## V. CONCLUSION

A tracking control method was proposed for switched systems with minimal number of mode switches. First the problem was solved for a hybrid system and then the solution for switched systems was presented by restricting the switched system into a hybrid system. Solvability conditions were also provided and it was shown that how the output can be modified to make the problem solvable. Application to control of induction motor shows a superior performance.